\documentclass[twocolumn,prb]{revtex4}
\usepackage{graphicx}
\usepackage{dcolumn}
\usepackage{bm}

\usepackage{graphics}

\begin{document}

\title{A simple derivation of the Gompertz law for human mortality.}

\author{B.~I.~Shklovskii}

\affiliation{William I. Fine Theoretical Physics Institute,
University of Minnesota, Minneapolis, Minnesota 55455}

\date{\today}
%%%%%%%%%%%%%%
\begin{abstract}

The Gompertz law for the human mortality rate as a function of age
is derived from a simple model of death as a result of an
exponentially rare escape of abnormal cells from the immunological
response.

%%%%%%%%%%%%%%%%%%%%%
\end{abstract}

\maketitle

Human life is finite as are the lives of some well known physical
objects for example radioactive nuclei. The probability, $S(t)$,
that a given nucleus will survive time $t$ is
\begin{equation}
S(t)= \exp(-t/t_0). \label{nuclei}
\end{equation}
If survival of humans were governed by the same law with, for
example, $t_0 =70$ years we would have millions of people with the
age, say, $5t_0= 350$ years. However, the longest recorded human
life lasted only 122 years. Apparently, the law of human mortality
should be drastically different from Eq.~(\ref{nuclei}).

The famous astronomer Halley~\cite{Halley} and the great
mathematician Euler~\cite{Euler} were the first to attack this
problem, but the law was empirically found only in 1825 by the
actuary and self-taught mathematician Benjamin
Gompertz~\cite{Gompertz}. If $S(t)$ is probability of a human
surviving till age $t$ years then the mortality function $\mu(t)$
is defined as
\begin{equation}
\mu(t) = -d\ln S(t)/dt. \label{mortfun}
\end{equation}
Gompertz found that when childhood diseases are overcome ($t >
25$) the statistics of mortality obeys the law
\begin{equation}
\mu(t) = \mu(25)\exp([t-25]/t_{1}), \label{Gom}
\end{equation}
where $t_{1}\simeq 10$. This law should be contrasted with the
constant mortality function $\mu(t) = 1/t_0$ following from
Eq.~(\ref{nuclei}). The Gompertz law means that $S(t)$ decays
double exponentially, much faster than Eq.~(\ref{nuclei}),
practically eliminating people older than 122 in the world
population. Statistical data for many countries and three
centuries confirm the Gompertz law in the range of $\mu(t)$
covering more than three orders of magnitude (see
Refs.~\cite{Stauffer} and references therein). The law works also
for different species such as rats, mice and fruit flies. Thus,
the Gompertz law emerges as one of the greatest quantitative laws
of biology.

There were a number of attempts to derive the Gompertz law. The
most quoted derivation uses the langauge of the reliability theory
designed for man-made machines~\cite{Gavrilov} and is quite
complicated. Here, we would like to suggest a simple version of
derivation of Eq.~(\ref{Gom}), based on a naive understanding of
immunology. Thinking about cells, we assume that a population of
defective cells (mutated cells that do not fulfill their normal
function, cancer cells undergoing uncontrolled proliferation,
cells that produce masses of defective protein forming amyloids
and senile plaques in aged tissues leading to diseases like
Alzheimer) becomes fatal if the normal organism defense system
gives this population a time $\tau$ necessary to reach a critical
size. Let us assume that at age 25, bad cells would encounter
strong and fatal immune response during time $\tau$ on average
$N_0 \gg 1$ times. (We can imagine a bad cell as criminal who is
being eliminated at any encounter with randomly patrolling
policemen (macrophages, natural killers, or apoptosis), before
growing strong enough to successfully resist them. Until this
time, on average, he would meet $N_0$ policemen.) For absolutely
random encounters, the probability $P(N)$ of the number $N$ of
encounters with immune response system during time $\tau$ is given
by the Poisson distribution,
\begin{equation}
P(N) = \frac{N_0^{N}\exp(-N_0)}{N!},
 \label{Poisson}
\end{equation}
where $N_0$ is the average value of $N$. The probability of a
population of defective cells reaching the critical size and
causing the death of the host at age 25 is $\mu(25) = P(0) =
\exp(-N_0)$. It is known that $\mu(25)$ is of the order of
$3~10^{-4}$. This means that, indeed, $N_0 \simeq 8 \gg 1 $. Let
us assume now that the immune response slowly weakens with the age
so that at $t> 25$ the average number of encounters during
"microscopic" time $\tau$ decreases linearly as $N_{0}(t)= N_0 -
(t-25)/t_{1}$. This may happen due to accumulation of mutations in
immune response cells, their limited potential for self-renewal,
or an overall decay of the organism energetics. (In the language
of criminals and policemen, this would mean slow decay of the
number of patrolling policemen, for example, due to budget
restrictions). Substituting this $N_{0}(t)$ instead of $N_0$ into
the Poisson formula for $P(0)$ we arrive at $\mu(t) = \exp[-
N_{0}(t)]$ and, therefore, at Eq.~(\ref{Gom}).

Notice that a relatively small change of $N_0(t)$ leads to a very
strong exponential growth of $\mu(t)$. A linear decay of
$N_{0}(t)$ can be the initial stage of an exponential relaxation
$N_{0}(t)= N_{0}\exp(-[t-25]/N_{0}t_{1})$. In this case, at large
ages $t\sim 100$ we arrive at downward deviations of $\mu(t)$ from
the Gompertz law. This qualitatively agrees with somewhat slower
than Eq.~(\ref{Gom}) growth of mortality at $t > 100$ cited in
~\cite{Azbel,Stauffer,Gavrilov,Alm}. These empirical deviations
from Eq.~(\ref{Gom}) are still under discussion because of
relatively poor statistics for $t > 100$.

I am grateful to M. Azbel, A. Finkelstein, A. Chklovski, D.
Chklovskii, S. Grigoryev, I. Ruzin, L. Shklovskii and D. Stauffer
for useful discussions.

%%%%%%%%%%%%%%%%%%%%%%%%%%%%%%%%%%%%%%%%%%%%%%%%%%%%%%%%%%%%%%%%

\end{document}